\documentclass[twocolumn,english,aps,pra,amsfonts,amsmath,amssymb,footinbib,floatfix,longbibliography]{revtex4-2}
\usepackage[latin9]{inputenc}
\usepackage{float}
\usepackage{dsfont}
\usepackage{amsmath}
\usepackage{amssymb}
\usepackage{graphicx}
\usepackage{esint}

\makeatletter
\usepackage{babel}
\usepackage{color}

\def\im{\mathrm{i}}

\newcommand{\hidetxt}[1]{}

\makeatother

\usepackage{babel}
\begin{document}
\title{Synthesizing arbitrary dispersion relations in a modulated tilted
optical lattice}
\author{Jean Claude Garreau}
\affiliation{Univ. Lille, CNRS, UMR 8523 - PhLAM - Laboratoire de Physique des
Lasers Atomes et Mol{\'e}cules, F-59000 Lille, France}
\author{V{\'e}ronique Zehnl{\'e}}
\affiliation{Univ. Lille, CNRS, UMR 8523 - PhLAM - Laboratoire de Physique des
Lasers Atomes et Mol{\'e}cules, F-59000 Lille, France}
\date{\today}
\begin{abstract}
Dispersion relations are fundamental characteristics of the dynamics
of quantum and wave systems. In this work we introduce a simple technique
to generate arbitrary dispersion relations in a modulated tilted lattice.
The technique is illustrated by important examples: the Dirac, Bogoliubov
and Landau dispersion relations (the latter exhibiting the roton and
the maxon). We show that adding a slow chirp to the lattice modulation
allows one to reconstruct the dispersion relation from dynamical quantities.
Finally, we generalize the technique to higher dimensions, and generate
graphene-like Dirac points and flat bands in two dimensions. 
\end{abstract}
\maketitle

\section{Introduction }

Dispersion relations, connecting the energy to the momentum, $E=E(\boldsymbol{p})$,
of a quantum particle, or the frequency to the wave number $\omega=\omega(\boldsymbol{k})$
of a wave, are a fundamental concept in many domains of physics. For
example, relativistic particles are characterized by the Einstein's
dispersion relation $E^{2}=p^{2}c^{2}+m^{2}c^{4}$, crystalline solids
by their bands $E=E(\boldsymbol{q})$ (with $\boldsymbol{q}$ the
quasimomentum) and superfluidity by Landau's dispersion relation presenting
exotic features like the \emph{maxon} and the\emph{ roton}~\citep{Landau:TheorySuperfluidityHeII:PR41,Leggett:QuantumLiquids:06,Barenghi:PrimerQuantumFluids:arXiv16}.
Dispersion relations provide a great deal of information on the physics
of a system.

Recent developments both in condensed matter and ultracold-atom systems
have generated the concept of ``quantum simulator''. Devising, for
instance, a system displaying a given dispersion relation provides
information on several aspects of the physics of other systems exhibiting
the same dispersion relation. Ultracold atoms in optical lattices
or trapped ions have proved to be one of the most clean and flexible
systems in physics, both theoretically and experimentally. Optical
lattices can mimic an almost perfect lattice (no phonons, controllable
decoherence, etc.) and make it easy to create low-dimensional systems.
Moreover, being formed by the interference of laser beams, they can
be ``engineered'' in many ways, e.g one can create a wealth of different
lattices. A few, non-exaustive, examples are Kagome~\citep{Jo:UltracoldAtomsTunableOpticalKagomeLattice:PRL12},
Lieb~\citep{Slot:ExperimentalRealizationLiebLattice:NP17,Goldman:TopologicalPhasesLieb:PRA11,Flannigan:HubbardModelsLiebLattice:arXiv21},
quasiperiodic~\citep{Roati:AubryAndreBEC1D:N08} and disordered lattices~\citep{Billy:AndersonBEC1D:N08}.
They can also be easily modulated in time, producing, notably, oscillating~\citep{Moore:LDynFirst:PRL94}
or accelerated lattices~\citep{BenDahan:BlochOsc:PRL96,Raizen:WSOptPot:PRL96}.
These properties make such systems an outstanding platform to the
realization of analog quantum simulators. In many cases, one can construct
a given Hamiltonian from its building blocks, e.g. Bose- and Fermi-Hubbard's~\citep{Jaksch:ColdBosonicAtomsInOpticalLattices:PRL98,Greiner:MottTransition:N02,Trimborn:MeanFieldlBoseHubbard:PRA09},
Anderson's~\citep{Billy:AndersonBEC1D:N08,Roati:AubryAndreBEC1D:N08,Moore:AtomOpticsRealizationQKR:PRL95,Chabe:Anderson:PRL08}
and Dirac's~\citep{Gerritsma:QuantumSimulationDirac:N10,Witthaut:EffectiveDiracDynamicsBichrOptLatt:PRA11,Tarruell:MergingDiracPtsHoneycombLattice:N12,Garreau:SimulatingDiracModelsUltracoldBosos:PRA17,Garreau:QuantumSimulDiracSp4GaugeField:PRA20}
Hamiltonians, bringing new light into many aspects of their physics,
notably the Mott transition~\citep{Greiner:MottTransition:N02},
the Anderson localization and transition~\citep{Chabe:Anderson:PRL08,Kondov:ThreeDimensionalAnderson:S11,Jendrzejewski:AndersonLoc3D:NP12,Semeghini:MobilityEdgeAnderson:NP15},
Bloch oscillations and Wannier-Stark systems~\citep{BenDahan:BlochOsc:PRL96,Raizen:WSOptPot:PRL96},
or the Klein tunneling~\citep{Gerritsma:QuantumSimulationKleinParadox:PRL11,Suchet:AnalogSimulationWeylParticles:EPL16}. 

In the present work we present a simple one dimensional system able
to ``synthesize'' almost \emph{any} desired dispersion relation,
that we illustrate with important examples. An original technique
allowing the direct detection of these dispersion relations is also
proposed. In the last part, we generalize the synthesis of dispersion
relations to higher dimensions, illustrated by the generation of graphene-like
Dirac cones and Lieb lattices (displaying a flat band) in 2D. 

\section{Synthesizing dispersion relations in one-dimensional Wannier-Stark
optical lattices}

\label{sec:DispRelWS}

Our present model is based on the framework introduced in Ref.~\citep{Garreau:SimulatingDiracModelsUltracoldBosos:PRA17}
and further developed in~\citep{Garreau:QuantumSimulDiracSp4GaugeField:PRA20}.
Ultracold atoms are placed in the interference pattern of counter-propagating
laser beams which generates a sinusoidal potential, proportional to
the atom-laser coupling, acting on on the center-of-mass degree of
freedom of the atoms~\citep{Cohen-TannoudjiDGO:AdvancesInAtomicPhysics::11}.
The atomic cloud density is assumed low enough that atomic interactions
are negligible. We consider here a one-dimensional Wannier-Stark system
consisting of a sinusoidal optical lattice to which a constant force
$F$ is applied:
\[
H=\frac{p^{2}}{2m}-V_{0}\cos\left(\frac{2\pi x}{a}\right)+Fx.
\]
This Hamiltonian can obtained by placing ultracold atoms on an accelerated
standing laser wave~\citep{BenDahan:BlochOsc:PRL96,Raizen:WSOptPot:PRL96,Manai:Anderson2DKR:PRL15}~\footnote{The force can be generated by applying a linear chirp to one of the
beams forming the standing wave, so that the nodes of the resulting
standing wave are uniformly accelerated. In the (non-inertial) reference
frame where the standing wave is at rest, the atoms feel a constant
inertial force.} of wavelength $k_{L}$, hence $a=\pi/k_{L}$. The properties of such
a system are well known~\citep{Wannier:WannierStates:PR37,Korsch:LifetimeWS:PRL99,Glueck:WannierStarkLadderDrivenOptLatt:PRA00,Nienhuis:CoherentDyn:PRA01,Thommen:WannierStark:PRA02,Korsh:WSRev:PREP02,Ploetz:EffectiveSpinModelForInterband:21}.
In short, this system is invariant under discrete spatial translations
corresponding to an integer multiple $n$ of the lattice step $a$
provided the energy is also shifted of $nFa=n\hbar\omega_{B}$ where
$\omega_{B}\equiv Fa/\hbar$ is the so-called \emph{Bloch frequency}.
In the following, $n$ denotes the site index corresponding to potential
minima localized at the position $x=na$. The symmetry of the system
implies that the eigenfunctions (called Wannier-Stark states) are
invariant under a translation of an integer number of lattice steps,
i.e $\varphi_{n}^{(\ell)}(x)=\varphi_{0}^{(\ell)}(x-na)$, with the
corresponding eigenenergies $E_{n}^{(\ell)}=E_{0}^{(\ell)}+n\hbar\omega_{B}$,
thus forming (depending on parameters $V_{0}$ and $F$) different
ladders (labeled by $\ell$) of levels separated by a step $\hbar\omega_{B}$
and characterized by a ground energy shift $E_{0}^{(\ell)}$. In what
follows we will mainly consider potential parameters allowing two
ladders $(\ell=g,e)$ of localized ``ground'' and ``excited''
eigenstates, that is, there are two states localized at each site
$n$, $\varphi_{n}^{(g)}(x)$ and $\varphi_{n}^{(e)}(x)$, separated
by an energy shift $\Delta=E_{0}^{(e)}-E_{0}^{(g)}$. Other eigenstates
of $H$, localized or belonging to the continuum, are supposed to
be irrelevant for the system's dynamics, as explained below. 

A convenient set of dimensionless units is obtained by measuring space
in units of $a$, energy in units of the so-called atom's \emph{recoil
energy} $E_{R}\equiv\hbar^{2}k_{L}^{2}/2m$$=\hbar\omega_{R}$ and
time in units of $\omega_{R}^{-1}$ ($m$ is the atom's mass). This
leads to the dimensionless Hamiltonian
\begin{equation}
H_{0}=\frac{p^{2}}{2m^{*}}-V_{0}\cos\left(2\pi x\right)+\omega_{B}x\label{eq:H0}
\end{equation}
where $m^{*}=\pi^{2}/2$, $\omega_{B}$ is measured in $\omega_{R}$
units and Planck's constant is $\hbar=1$. 

Controllable dynamics can be induced in such systems by external modulations
of the parameters $V_{0}$ (or $F$) at frequencies close to resonances,
i.e close to $\Delta$ and to multiples of $\omega_{B}$. We thus
add to $H_{0}$ a time-dependent potential
\begin{equation}
H_{1}(t)=f(t)V(x)\label{eq:H1}
\end{equation}
where $V(x)=\cos(2\pi x)$. Using this flexibility, the feasibility
of models reproducing the Dirac equation (hence the Dirac dispersion
relation) has been demonstrated in Refs.~\citep{Garreau:SimulatingDiracModelsUltracoldBosos:PRA17,Garreau:QuantumSimulDiracSp4GaugeField:PRA20}.

The driving $f(t)$ has the general form
\begin{equation}
f(t)=\sum_{j,q}A_{j,q}\mathrm{e}^{\im j\omega_{B}t}\mathrm{e}^{\im q\Delta t}\label{eq:modulationGeneral}
\end{equation}
with $j\in\mathds{Z}$, $q=0,\pm1$ and $A_{j,q}=-A_{-j-q}^{*}$ (reality
condition). For given integers $j$ the modulation resonantly couples
states centered in sites separated by $j$ lattice steps. If $q=0$,
one couples states $\varphi_{n}^{(\ell)}$ and $\varphi_{n+j}^{(\ell)}$
belonging to the same ladder $\ell$ (\emph{intraladder} coupling).
\emph{Interladder} couplings between states $\varphi_{n}^{(g)}$ and
$\varphi_{n+j}^{(e)}$ are obtained for $q=1$, or $\varphi_{n}^{(e)}$
and $\varphi_{n+j}^{(g)}$ for $q=-1$. 

With the above provisions, the general solution for the system can
be developed in the Wannier-Stark basis restricted to the subspace
spanned by the ground and first excited ladder 
\begin{align}
\Psi(x,t)= & \sum_{n}\left(c_{n}(t)\mathrm{e}^{-\im n\omega_{B}t}\varphi_{n}^{(g)}(x)\right.\nonumber \\
 & \left.+d_{n}(t)\mathrm{e}^{-\im\left(n\omega_{B}+\Delta\right)t}\varphi_{n}^{(e)}(x)\right)\label{eq:two-ladders_expansion}
\end{align}
which holds if modulation amplitudes in $f(t)$ are low enough to
avoid projection on other eigenstates of the Hamiltonian $H_{0}$.

Plugging this form into the Schr\"odinger equation for $H_{0}+H_{1}(t)$
one obtains a set of coupled differential equations for the amplitudes
$c_{n}(t)$ and $d_{n}(t)$ (for details of this calculation, see
the Appendix of Ref.~\citep{Garreau:SimulatingDiracModelsUltracoldBosos:PRA17}),
which reads
\begin{align}
\im & \frac{d}{dt}\left(\begin{array}{c}
c_{n}(t)\\
d_{n}(t)
\end{array}\right)=\sum_{r\in\mathds{Z}}\left(\begin{array}{cc}
T_{r}^{(gg)} & T_{r}^{(ge)}\\
T_{r}^{(eg)} & T_{r}^{(ee)}
\end{array}\right)\left(\begin{array}{c}
c_{n+r}(t)\\
d_{n+r}(t)
\end{array}\right)\label{eq:gendiscretemodel}
\end{align}
with coupling amplitudes between sites $n$ and $n+r$~\footnote{The translation symmetry $\varphi_{n}^{(\ell)}(x-na)=\varphi_{0}^{(\ell)}(x)$
implies $\left\langle \varphi_{n}^{(\ell)}\right|V\left|\varphi_{n+r}^{(\ell)}\right\rangle =\left\langle \varphi_{0}^{(\ell)}\right|V\left|\varphi_{r}^{(\ell)}\right\rangle $for
$V=\cos(2\pi x)$.}
\begin{align}
T_{r}^{(gg)} & =A_{r,0}\left\langle \varphi_{0}^{(g)}\right|V\left|\varphi_{r}^{(g)}\right\rangle \nonumber \\
T_{r}^{(ee)} & =A_{r,0}\left\langle \varphi_{0}^{(e)}\right|V\left|\varphi_{r}^{(e)}\right\rangle \nonumber \\
T_{r}^{(ge)} & =A_{r,1}\left\langle \varphi_{0}^{(g)}\right|V\left|\varphi_{r}^{(e)}\right\rangle \nonumber \\
T_{r}^{(eg)} & =A_{r,-1}\left\langle \varphi_{0}^{(e)}\right|V\left|\varphi_{r}^{(g)}\right\rangle .\label{eq:Tr}
\end{align}
In the momentum representation the corresponding amplitudes $\tilde{c}(k,t)=\sum_{n}c_{n}(t)\exp\left(-\im kn\right)$
(with analogous expressions for $d_{n}$) are governed by only \emph{two}
coupled equations
\begin{align}
\im & \frac{d}{dt}\left(\begin{array}{c}
\tilde{c}(k,t)\\
\tilde{d}(k,t)
\end{array}\right)=\left(\begin{array}{cc}
F_{g}(k) & F(k)\\
F^{*}(k) & F_{e}(k)
\end{array}\right)\left(\begin{array}{c}
\tilde{c}(k,t)\\
\tilde{d}(k,t)
\end{array}\right)\label{eq:gendiscretemodel-kspace}
\end{align}
where
\begin{equation}
F_{\ell}(k)=\sum_{r=-\infty}^{\infty}T_{r}^{(\ell\ell)}\mathrm{e}^{\im kr}\label{eq:Fa}
\end{equation}
with $\ell=\left\{ g,e\right\} $, and 
\begin{equation}
F(k)=\sum_{r=-\infty}^{\infty}T_{r}^{(ge)}\mathrm{e}^{\im kr}.\label{eq:F(k)}
\end{equation}
The functions $F_{\ell}(k)$ and $F(k)$ are Fourier series whose
coefficients are proportional to the overlap integrals $T_{r}^{\ell\ell^{\prime}}$
of the wave functions centered at positions separated by $r$ sites
but \emph{are controlled by the modulation coefficients} $A_{j,q}$
{[}see Eqs.~(\ref{eq:Tr}){]}.

Equations~(\ref{eq:gendiscretemodel-kspace}) define an effective
Hamiltonian for a two-level-like system, whose eigenvalues determine
the dispersion relation $\omega(k)$:
\begin{equation}
\left[\omega(k)-F_{g}(k)\right]\left[\omega(k)-F_{e}(k)\right]=\left|F(k)\right|^{2}.\label{eq:dispGeneral}
\end{equation}
This dispersion relation is $2\pi$-periodic and is conventionally
defined in the interval $k\in\left[-\pi,\pi\right)$. Equation~(\ref{eq:dispGeneral})
is the basis of our technique for synthesizing dispersion relations:
Specifying a given dispersion relation amounts to define the functions
$F_{g}(k)$, $F_{e}(k)$ and $F(k)$ and, from Eqs.~(\ref{eq:Fa},\ref{eq:F(k)}),
to set conditions on the amplitudes $T_{r}^{(\ell,\ell^{\prime})}$,
which allows one, with the help of Eqs.~(\ref{eq:Tr}), to determine
modulation parameters $A_{j,q}$ that must be used in the modulation
term $H_{1}(t)$. Examples discussed in Sec.~\ref{sec:Applications}
illustrate how a desired dispersion relation can be obtained to a
very good approximation. 

\section{Applications}

\label{sec:Applications}

\subsection{Dirac cone}

A linear dispersion relation $\omega(k)=\pm c|k|$, also called Dirac
cone (corresponding to a Dirac particle of zero mass), can be obtained
from Eq.~(\ref{eq:dispGeneral}) with $F_{g}(k)=F_{e}(k)=0$ and
$F(k)=c\left|k\right|$. This implies $T_{r}^{(g,g)}=T_{r}^{(e,e)}=0$,
that is, no intraladder couplings, and interladder couplings given
by Eq.~(\ref{eq:F(k)}). As the Fourier series of the function $|k|$
is
\[
|k|=\left[\frac{\pi}{2}-\frac{4}{\pi}\sum_{r=1}^{\infty}\frac{\cos(rk)}{r^{2}}\right]
\]
one must have
\[
\sum_{r}T_{r}^{(g,e)}e^{\im kr}=c\left|k\right|=c\left[\frac{\pi}{2}-\frac{4}{\pi}\sum_{r}\frac{\cos(rk)}{r^{2}}\right]
\]
which, from Eq.~(\ref{eq:Tr}), gives
\begin{equation}
A_{0,1}\left\langle \varphi_{0}^{(g)}\right|V\left|\varphi_{0}^{(e)}\right\rangle =c\frac{\pi}{2}\label{eq:A01}
\end{equation}
 and
\[
A_{r,1}\left\langle \varphi_{0}^{(g)}\right|V\left|\varphi_{r}^{(e)}\right\rangle =\frac{c}{\pi r^{2}}\left[\cos(r\pi)-1\right]\qquad r>0.
\]

\begin{figure}[H]
\begin{centering}
\includegraphics[width=6cm]{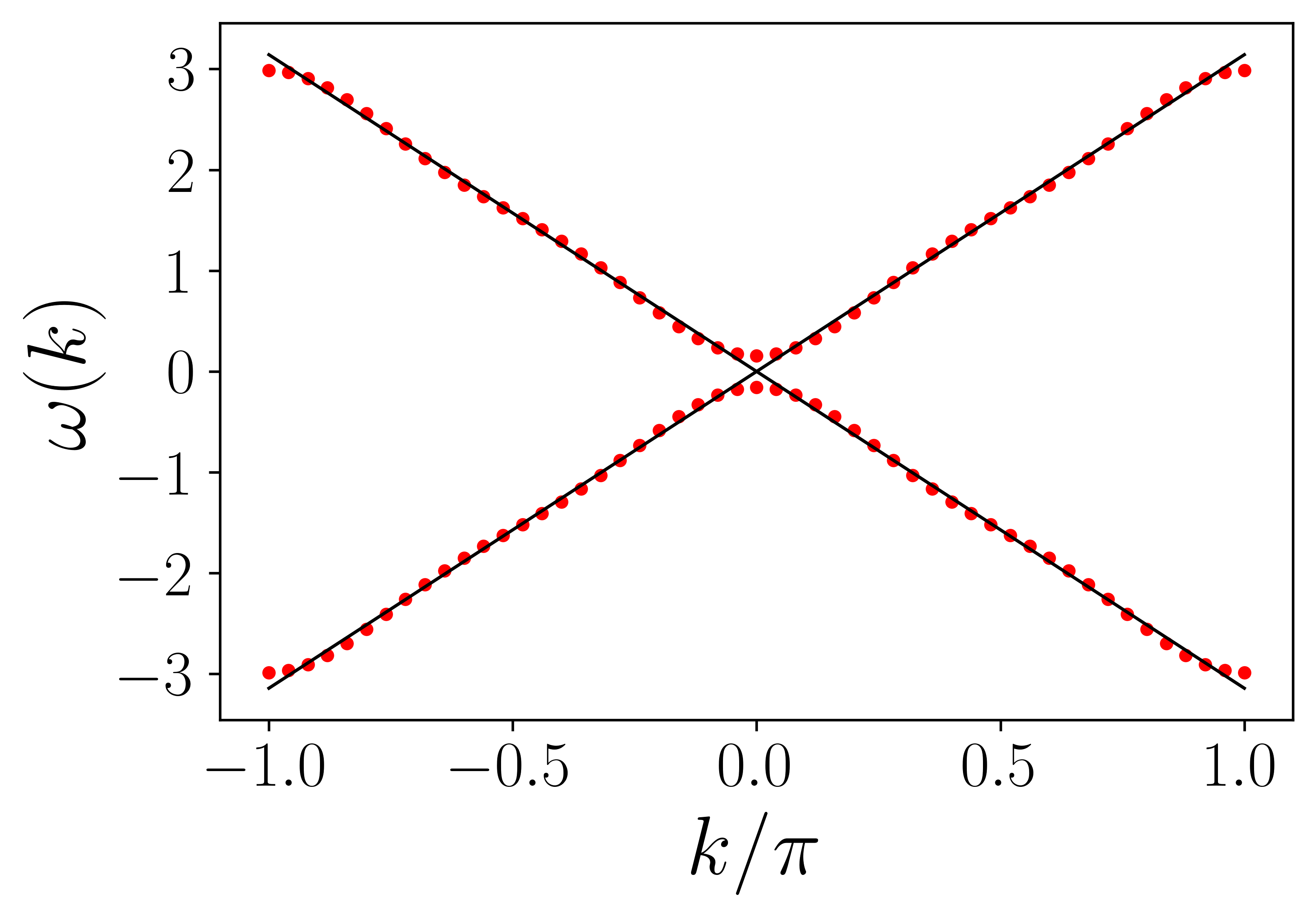}
\par\end{centering}
\begin{centering}
\includegraphics[width=6cm]{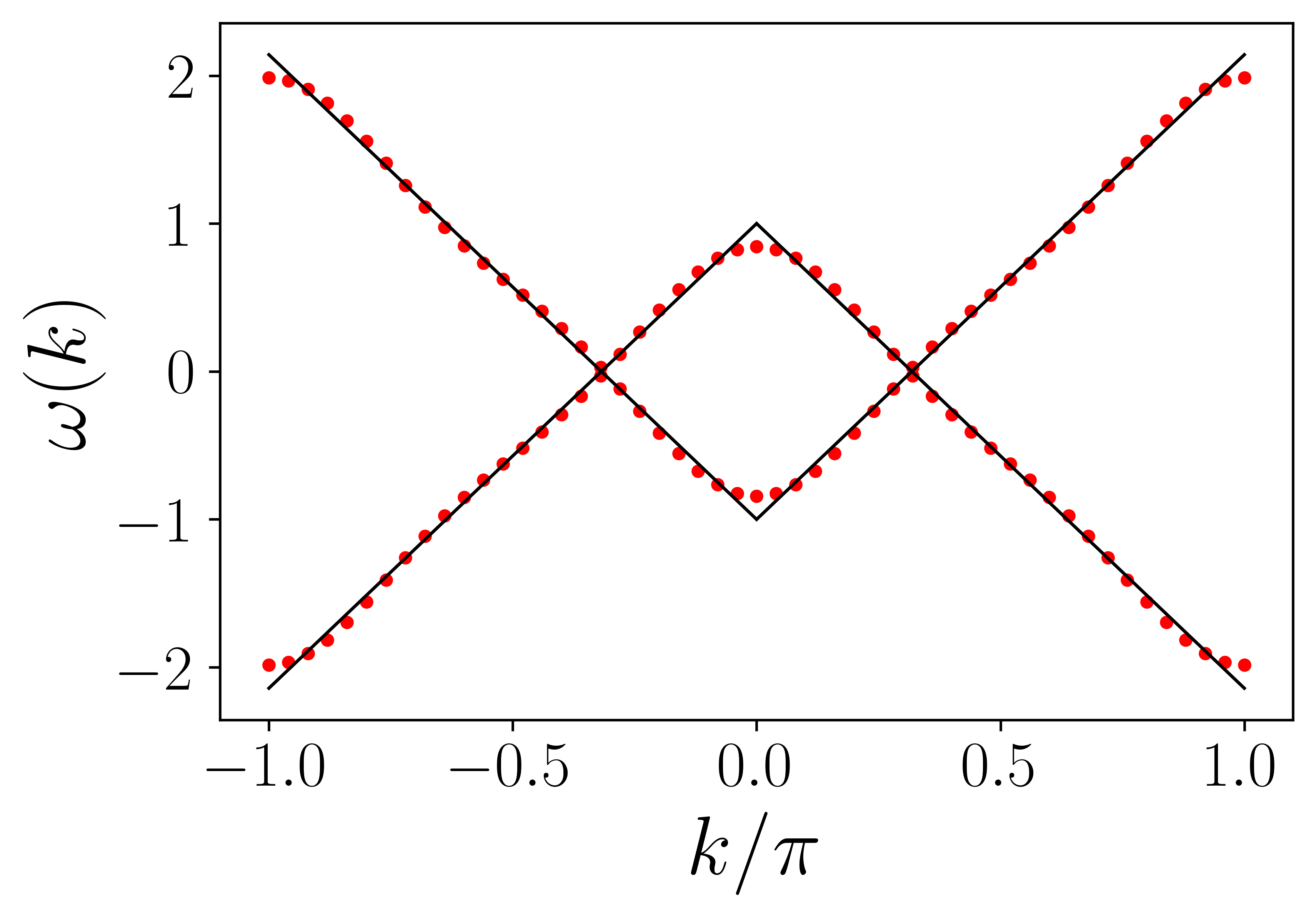}
\par\end{centering}
\caption{\label{fig:Linear}Linear dispersion relation. \emph{Top}: Dirac cone
($c=1$); \emph{Bottom}: intersecting Dirac cones ($c=1,b=-1$). Only
terms $0\le r\le3$ were kept in the Fourier series (red circles).
The analytical expressions are shown for comparison (black solid lines).}
\end{figure}

The dispersion relation is thus synthesized by applying a modulation
given by Eq.~(\ref{eq:modulationGeneral}) with coefficients:
\begin{align*}
A_{r,1}= & \frac{c\pi}{2}\left\langle \varphi_{0}^{(g)}\right|V\left|\varphi_{0}^{(e)}\right\rangle ^{-1}\delta_{r,0}\\
 & +\frac{c}{\pi r^{2}}\left\{ \cos(r\pi)-1\right\} \left\langle \varphi_{0}^{(g)}\right|V\left|\varphi_{r}^{(e)}\right\rangle ^{-1}\left(1-\delta_{r0}\right).
\end{align*}
Experimentally, this kind of modulation can be easily created by an
arbitrary-wave generator. The overlap integrals $\left\langle \varphi_{0}^{(g)}\right|V\left|\varphi_{r}^{(e)}\right\rangle $
depend on the lattice parameters $V_{0}$ and $F$ and go to zero
rather fast with the site distance $r$, implying that the modulation
amplitude at higher frequencies must increase rapidly, eventually
breaking the slow modulation condition implied by Eq.~(\ref{eq:two-ladders_expansion}).
Fortunately, for typical lattice parameters, keeping only $r\le3$
terms yet gives a rather good approximation of a Dirac cone, as shown
Fig.~\ref{fig:Linear} (\emph{top}), except at the tip of the cone
at $k=0$ which needs higher harmonics to be well reproduced. 

A way to overcome this limitation is to include in Eq.~(\ref{eq:A01})
a frequency offset parameter $b$ such that $\omega(k)=\pm c\left(|k|+b\right)$,
which is readily done by setting
\[
A_{0,1}=c\left(b+\frac{\pi}{2}\right)\left\langle \varphi_{0}^{(g)}\right|V\left|\varphi_{0}^{(g)}\right\rangle ^{-1}.
\]
The parameter $b$ controls the gap between the two half-cones: If
$b>0$ a gap is opened, and if $b<0$ one obtains two intersecting
Dirac cones. Figure~\ref{fig:Linear} (\emph{bottom}) illustrates
this case for $b=-1$ and shows perfect linear behavior in the vicinity
of the Dirac points whose position $k=\pm b$ is controlled by the
applied modulation. We have thus obtained flexible way to generate
Dirac-cone-type dispersion relations, allowing one to open a gap or
to merge the cones.

\subsection{The superfluid dispersion relation, the roton and the maxon}

Superfluidity is a purely quantum behavior appearing notably in liquid
Helium and quantum atomic gases. Its simplest mathematical treatment
relies on the Bogoliubov correction to the mean-field Gross-Pitaevskii
equation~\citep{PethickSmith:BoseEinstein:08,Cohen-TannoudjiDGO:AdvancesInAtomicPhysics::11,Barenghi:PrimerQuantumFluids:arXiv16},
which leads to the well-known Bogoliubov dispersion relation
\begin{equation}
\omega(k)=\sqrt{\frac{k^{2}}{2M}\left(\frac{\hbar^{2}k^{2}}{2M}+2gn\right)}\label{eq:Bogo}
\end{equation}
where $M$ is the atom mass, $g$ a parameter characterizing the binary
atomic interaction in the Gross-Pitaevskii equation and $n$ the atomic
density (the approximation is valid in the limit of low density and
temperature). This dispersion is characterized by a ``phononic'',
i.e. linear $\omega(k)\propto k$ dependence at small $k$ $(k\ll\sqrt{gnM}/\hbar$)
and a ``free particle'' part $\omega(k)\propto k^{2}$ at large
$k$. 

Following the same ideas as in the preceding section, Eq.~(\ref{eq:Bogo})
can be obtained from the general form of the dispersion relation with
$F_{g}(k)=F_{e}(k)=0$ and 
\begin{equation}
F(k)=c\left(\left|k\right|+\alpha k^{2}\right)\label{eq:bogoliubov_roton}
\end{equation}
where $\alpha$ is an imaginary number ($\alpha=\im|\alpha|$), leading,
through Eq.~(\ref{eq:dispGeneral}), to the desired dispersion
\begin{equation}
\omega(k)=c\sqrt{k^{2}\left(1+\left|\alpha\right|^{2}k^{2}\right)}\label{eq:bogo}
\end{equation}

Beyond the Bogoliubov approach, phenomenological arguments by Landau~\citep{Landau:TheorySuperfluidityHeII:PR41}
concerning the superfluidity of the strongly-interacting liquid $^{4}$He
supported the existence in the dispersion relation of a minimum $\omega(k_{r})$
at some $k_{r}$, called the \emph{roton}~\footnote{The roton minimum determines Landau's critical velocity given by slope
of the dispersion relation, thus setting a gap responsible for the
superfluidity.} which, by continuity, implies the existence of a maximum $\omega(k_{m})$
with $k_{m}<k_{r}$, called the \emph{maxon}. Roton and maxon features
have recently been observed experimentally using a Bose-Einstein condensate
in a modulated flat lattice~\citep{Ha:RotonMaxonExcitationBEC:PRL15},
in Erbium condensates with interactions controlled by Feshbach resonances~\citep{Chomaz:ObservationDipolarQG:NP18,Santos:RotonMaxonBEC:PRL03}
and in acoustic metamaterials~\citep{Chen:RotonLikeAcousticalDispersion:NCM21}.

\begin{figure}[H]
\begin{centering}
\includegraphics[width=6cm]{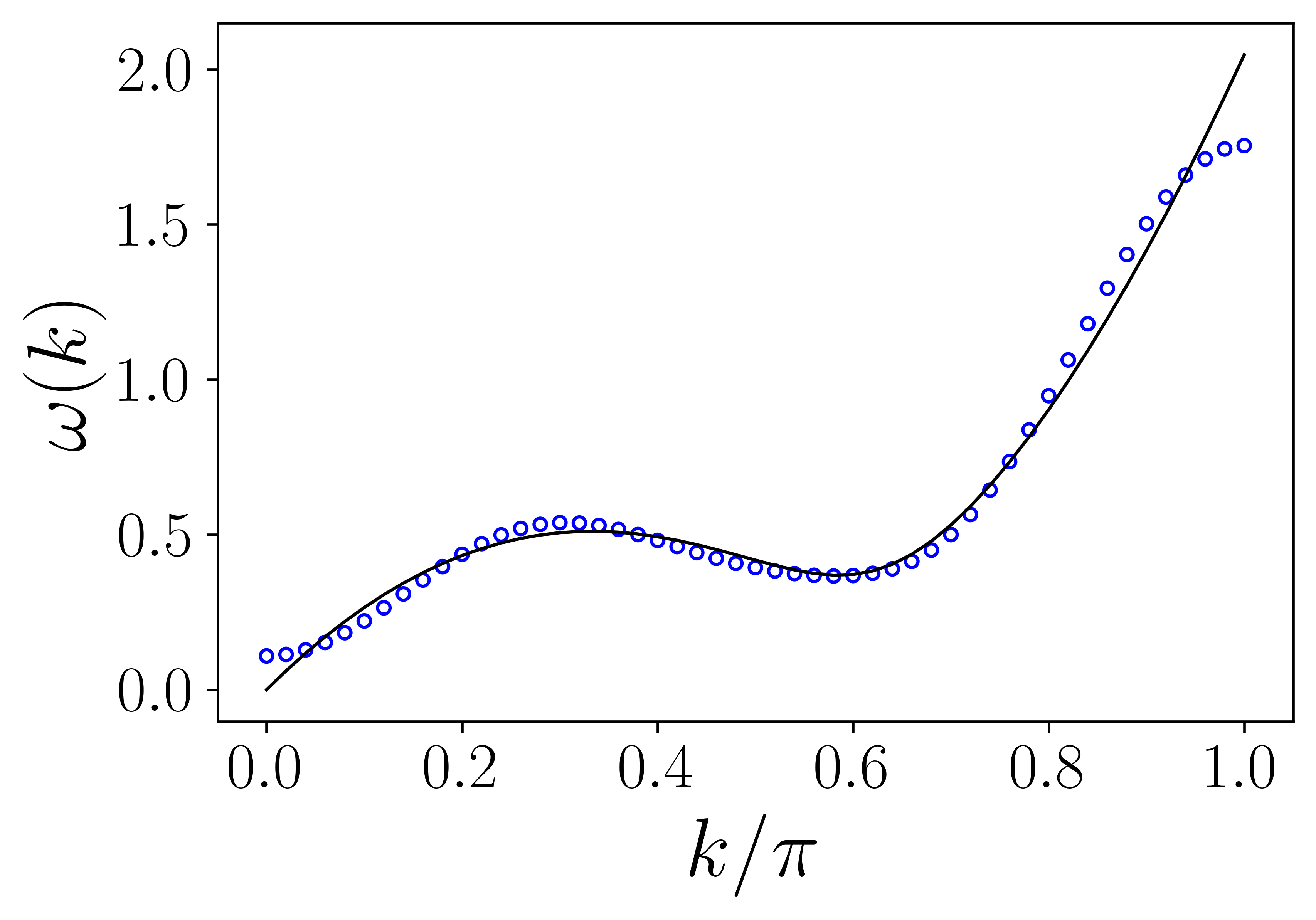}
\par\end{centering}
\caption{\label{Fig:Landau-roton}Synthetic dispersion relation displaying
a maxon and a roton, generated by the Fourier series Eq.~(\ref{eq:rotonfourierampl})
(blue circles), restricted to $|r|\le3$, compared to Eq.~(\ref{eq:stddisprelation})
with parameters $c=1$ and $\alpha=-0.5+0.1\im$ (black solid line).}
\end{figure}

A dispersion relation presenting a roton and maxon can be synthesized
by choosing $\alpha$ as a \emph{complex} number, thus
\begin{equation}
\omega(k)=c\sqrt{k^{2}\left(1+\left(\alpha+\alpha^{*}\right)|k|+\left|\alpha\right|^{2}k^{2}\right)}.\label{eq:stddisprelation}
\end{equation}
Equation~(\ref{eq:F(k)}) then leads to
\[
T_{r}^{ge}=A_{r,1}\left\langle \varphi_{0}^{g}\right|V\left|\varphi_{r}^{e}\right\rangle =\frac{c}{\pi}\intop_{0}^{\pi}\cos\left(rk\right)\left(\left|k\right|+\alpha k^{2}\right),
\]
from which we obtain the modulation amplitudes\begin{widetext}
\begin{equation}
A_{r,1}=c\left(\frac{\pi}{2}+\alpha\frac{\pi^{2}}{3}\right)\left\langle \varphi_{0}^{(g)}\right|V\left|\varphi_{0}^{(e)}\right\rangle ^{-1}\delta_{r0}+\frac{c}{\pi r^{2}}\left[(2\pi\alpha+1)\cos(r\pi)-1\right]\left\langle \varphi_{0}^{(g)}\right|V\left|\varphi_{r}^{(e)}\right\rangle ^{-1}\left(1-\delta_{r0}\right).\label{eq:rotonfourierampl}
\end{equation}
\end{widetext}The fact that $\left|A_{r,1}\right|\propto r^{-2}$
(for $r\neq0$) implies that the series has a rather fast convergence,
and thus only a few terms suffice to give a good approximation of
the desired dispersion relation. Figure~(\ref{Fig:Landau-roton})
shows the resulting Bogoliubov dispersion relation with the \textit{roton-maxon}
features obtained for $\alpha=-0.5+0.1\im$, compared to Eq.~(\ref{eq:stddisprelation}).

\subsection{Scanning the dispersion relation}

\label{subsec:Scanning-the-dispersion}

A natural question is, how to obtain experimental information on the
shape of the dispersion relation? We show below that this can be done
by preparing an initial wave packet and monitoring its average position
$\left\langle X\right\rangle (t)$, in the modulated lattice while
applying a slow phase chirp to the modulation Eq.~(\ref{eq:modulationGeneral});
we thus use a generalized modulation form $\sum_{j}A_{j,q}\exp\left(\im\left[j\left(\omega_{B}t+\phi(t)\right)+q\Delta t\right]\right)$,
where $\phi(t)$ is an arbitrary phase modulation. All the developments
of Sec.~\ref{sec:DispRelWS} remain valid except that, for a slow
variation of $\phi(t)$, the amplitudes become $A_{j,0}\exp\left(\im j\phi(t)\right).$
Equations~(\ref{eq:Tr}) show that phase modulation $\phi(t)$ is
then imprinted in the coupling coefficients $T_{r}$, which finally
results in shifting $k\rightarrow k(t)=k+\phi(t)$ in Eqs.~(\ref{eq:Fa},\ref{eq:F(k)}).
Changing the phase $\phi$ turns out to be equivalent to change $k$~\footnote{That is, $\phi$ acts as a (time-dependent) quasimomentum.},
resulting in slowly varying functions $F(k(t))$ and $F_{\ell}\left(k(t)\right)$
and thus $\omega\left(k(t)\right)$. 

\begin{figure}
\centering{}\includegraphics[width=6cm]{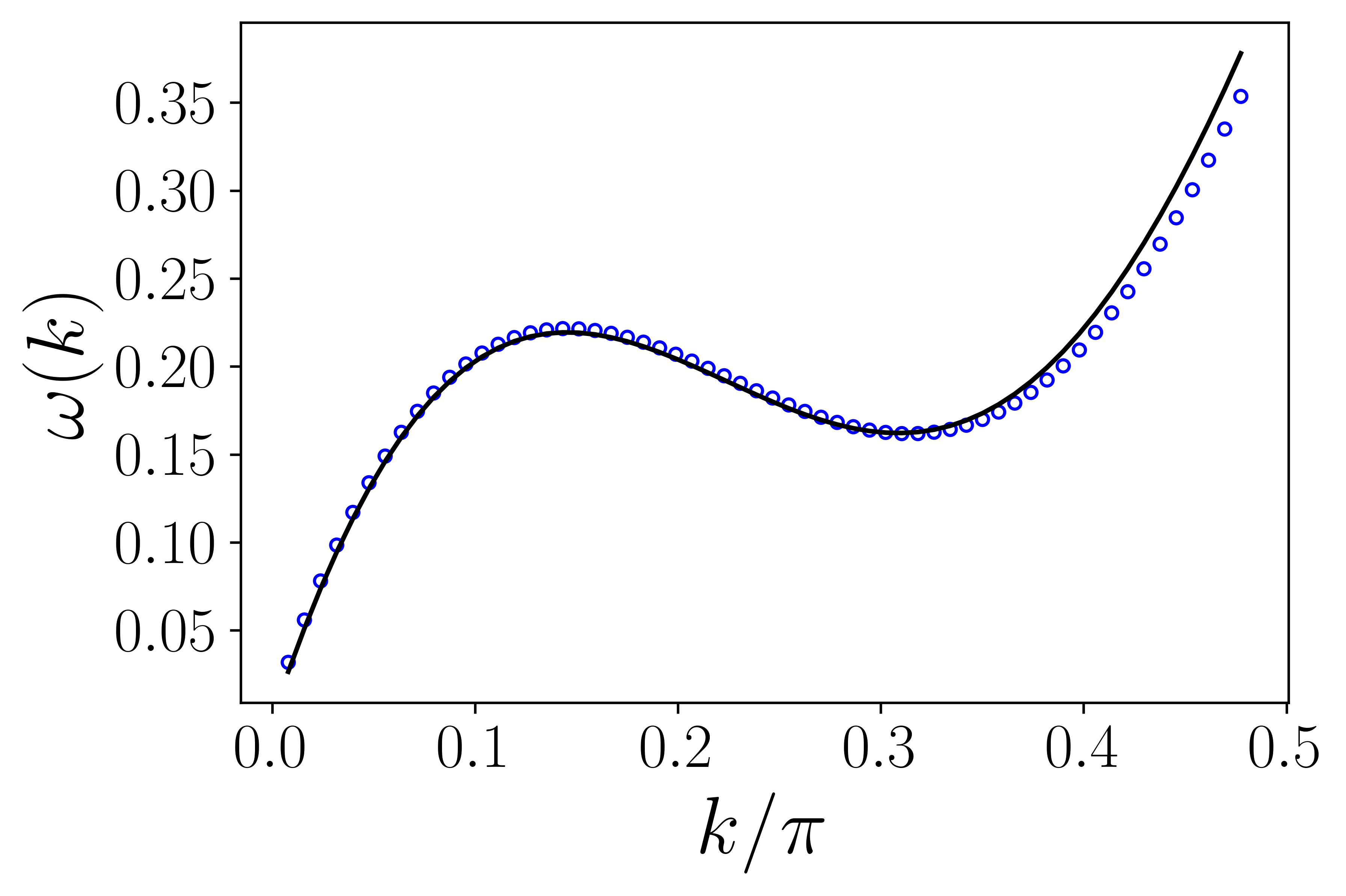}
\caption{\label{fig:num_sim}Detecting the dispersion relation. The solid black
line dispersion relation is obtained from $\omega(k)=F_{g}\left[k+\phi(t)\right]$
with $F_{g}(k)=-0.81\left[1-\cos(2k)\right]-0.5\sin(2k)+2\sin k$
and $F_{e}=F=0$. The blue circles show the average wave packet position
$\left\langle X\right\rangle (t)$ (with an adjusted vertical scale
factor) and the chirp ratio $\gamma=\pi\times10^{-4}\omega_{B}$. }
\end{figure}

We illustrate this idea in the simple case of an intraladder model
with only the ground ladder $\ell=0$ {[}i.e. $q=0$ in Eq.~(\ref{eq:H1}){]}.
In this case we have $\omega(k)=F_{g}\left[k+\phi(t)\right]$, and
a slow linear chirp $\phi=\gamma t$ ($\gamma\ll\omega_{B}$) results
in $\left\langle X\right\rangle (t)=\left.\omega(k)\right|_{k=k(t)}/\gamma$
(see appendix~\ref{app:Appendix chirp}). We choose here the example
of an analytical dispersion relation
\begin{equation}
\omega(k)=a\left[1-\cos(2k)\right]+b\sin(2k)+c\sin k\label{eq:DpRel}
\end{equation}
which exhibits a roton-like behavior as shown in Fig.~\ref{fig:num_sim}
(full black line). We performed a numerical simulation of the full
Schr\"odinger equation corresponding to the Hamiltonian $H=H_{0}+H_{1}(t)$
(with a suitable choice of modulation parameters) with an initial
Gaussian wave packet whose width in $k-$space is small compared to
the extent of the dispersion relation (or more precisely to its typical
length scale). The time evolution of the wave packet is then registered
while the phase is linearly swept $\phi(t)=\gamma t$. The evolution
of the mean position $\left\langle X\right\rangle (t)$ is shown in
Fig.~\ref{fig:num_sim} (blue circles) and shows a good agreement
with the analytical expression Eq.~(\ref{eq:DpRel}), validating
our proposal.

\section{Dispersion relations in two dimensions}

\label{sec:Disp_2D}

The strategy developed in Sec.~\ref{sec:DispRelWS} can be generalized
to higher dimensions. We give here two examples: the creation of 2D
Dirac points and of a Lieb lattice dispersion relation with a flat
band.

\subsection{Creating, moving and the merging of Dirac cones }

\label{subsec:2DDiracCones}

Manipulation of Dirac points both in optical lattices~\citep{Tarruell:MergingDiracPtsHoneycombLattice:N12}
and crystals~\citep{Montambaux:MergingOfDiracPts2DCrystal:PRB09}
is an active research subject. The creation of a 2D optical lattice
simulating Dirac physics was discussed in Ref.~\citep{Garreau:QuantumSimulDiracSp4GaugeField:PRA20}.
In brief, the corresponding dispersion relations can be synthesized
in a 2D modulated tilted lattice. Hamiltonian~(\ref{eq:H1}) is easily
translated into 2D (the definition of the dimensionless units is analogous
to the 1D case):

\begin{equation}
H_{0_{2D}}=\frac{p_{x}^{2}+p_{y}^{2}}{2m^{*}}-V_{0}\left(\cos(2\pi x)+\cos(2\pi y)\right)+\omega_{B}^{(x)}x+\omega_{B}^{(y)}y.\label{eq:H2}
\end{equation}
Since the above Hamiltonian is separable, all Wannier-Stark properties
described in Sec.~\ref{sec:DispRelWS} trivially generalize to the
present case. We consider only \textsl{ground} ladder eigenstates
formed by the tensorial product $\varphi_{n}^{(x)}(x)\varphi_{m}^{(y)}(y)$
with $\varphi_{s}^{(i)}(x_{i})=\varphi_{0}^{(i)}(x_{i}-sx_{0_{i}})$,
$i=\left\{ x,y\right\} ,s=\left\{ n,m\right\} $ with eigenenergies
$E_{0}+n\omega_{B}^{(x)}+m\omega_{B}^{(y)}$ .

The controlled dynamics for Dirac system we are interested in is generated
by adding a perturbation
\begin{align}
H_{1_{2D}}(x,y,t)= & \cos\left(\pi x\right)\left[V_{x}\mathrm{e}^{\im\left(2\omega_{B}^{(x)}t+\phi_{x}\right)}+\mathrm{c.c.}+V_{0}\right]\nonumber \\
+ & \left[V_{y}\cos\left(2\pi y\right)\mathrm{e}^{\im\left(\omega_{B}^{(y)}t+\phi_{y}\right)}+\mathrm{c.c.}\right]\label{eq:perturbation}
\end{align}
depending on the arbitrary phases $\phi_{x,y}$. The general solution
(restricted to the ground ladder) can be written, in analogy with
Eq.~(\ref{eq:two-ladders_expansion}),
\begin{equation}
\Psi(x,y,t)=\sum_{n,m}c_{n,m}(t)\varphi_{n}^{(x)}(x)\varphi_{m}^{(y)}(y).\label{eq:psi_2D}
\end{equation}
Substituting the above solution in Schrödinger's equation for the
total Hamiltonian $H_{0_{2D}}+H_{1_{2D}}$, one obtains the following
set of coupled equation (in the resonant approximation)
\begin{align*}
\im\frac{dc_{n,m}}{dt}= & (-1)^{n}\left[T_{x}\mathrm{e}^{\im\phi_{x}}c_{n+2,m}+T_{x}\mathrm{e}^{-\im\phi_{x}}c_{n-2,m}+T_{0}c_{n,m}\right]\\
 & +T_{y}\mathrm{e}^{\im\phi_{y}}c_{n,m+1}+T_{y}\mathrm{e}^{-\im\phi_{y}}c_{n,m-1}
\end{align*}
with couplings
\begin{align*}
T_{x} & =V_{x}\left\langle \varphi_{0}^{(x)}\right|\cos\left(\pi x\right)\left|\varphi_{2}^{(x)}\right\rangle \\
T_{y} & =V_{y}\left\langle \varphi_{0}^{(y)}\right|\cos\left(2\pi y\right)\left|\varphi_{1}^{(y)}\right\rangle \\
T_{0} & =V_{0}\left\langle \varphi_{0}^{(x)}\right|\cos\left(\pi x\right)\left|\varphi_{0}^{(x)}\right\rangle 
\end{align*}
where the factor $\cos(\pi x)$ in Eq.~(\ref{eq:perturbation}) introduces
a parity-dependent factor $(-1)^{n}$ which results in a system of
two coupled sublattices corresponding to sites where $n$ is odd or
even. In the reciprocal space $(k_{x},k_{y})$, we define two-dimensional
Fourier amplitudes, 
\[
\tilde{c}(k_{x},k_{y},t)=\sum_{n\textrm{ even}}\sum_{m}c_{nm}(t)\mathrm{e}^{-\im nk_{x}}\mathrm{e}^{-\im mk_{y}}
\]
for $n$ even, and, equivalently, for $n$ odd, $\tilde{d}(k_{x},k_{y},t)$.
These amplitudes can be written as a two-component spinor $\left[\psi\right]=\left(\tilde{c}(k_{x},k_{y},t),\tilde{d}(k_{x},k_{y},t)\right){}^{\mathsf{T}}$.
The Hamiltonian projected in the $k$-space turns out to be
\begin{equation}
\left[\begin{array}{cc}
T_{0}+T_{x}\cos(2k_{x}+\phi_{x}) & T_{y}\cos(k_{y}+\phi_{y})\\
T_{y}\cos(k_{y}+\phi_{y}) & -T_{0}-T_{x}\cos(2k_{x}+\phi_{x})
\end{array}\right]\label{eq:Dirac2D}
\end{equation}
for which the dispersion relation is
\begin{align}
\omega(k_{x},k_{y})= & \pm\left\{ \left[T_{0}+T_{x}\cos(2k_{x}+\phi_{x})\right]^{2}\right.\nonumber \\
 & \left.+\left[T_{y}\cos(k_{y}+\phi_{y})\right]^{2}\right\} ^{1/2}.\label{eq:disp_2D}
\end{align}
In the $k_{x,y}\rightarrow0$ limit, choosing $\phi_{x}=\phi_{y}=\pi/2$
and $T_{0}=0$, a Dirac equation (for a free particle) is obtained
with the dispersion relation 
\[
\omega(k_{x},k_{y})=\pm\sqrt{4T_{x}^{2}k_{x}^{2}+T_{y}^{2}k_{y}^{2}}
\]
which is a Dirac cone (anisotropic if $2T_{x}\neq T_{y}$). 

\begin{figure}
\begin{centering}
\includegraphics[width=6cm]{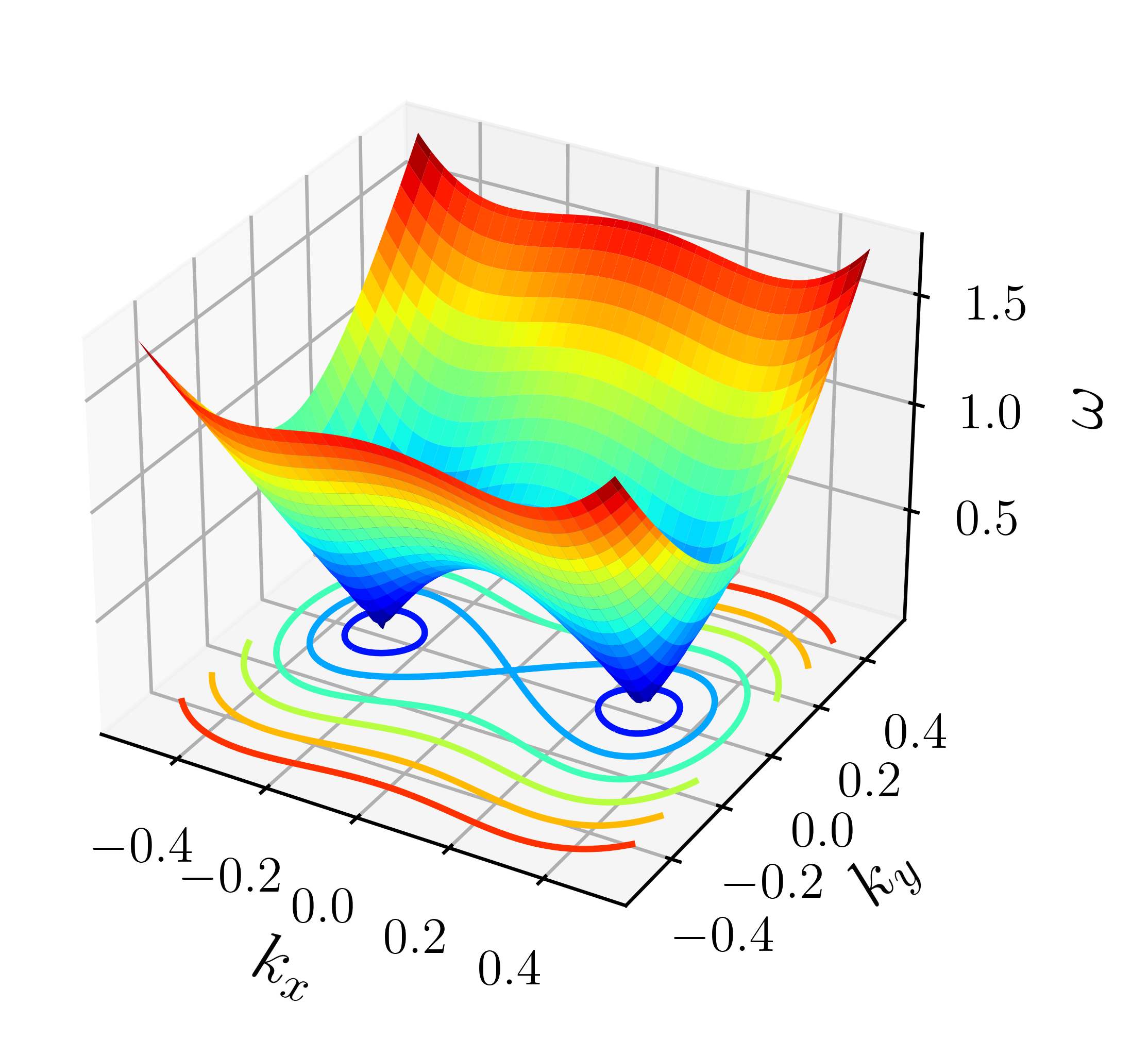}\\
\par\end{centering}
\caption{\label{fig:2D-Dirac}Dirac cones. We plot the positive root of Eq.~(\ref{eq:Two_Dirac})
for parameters: $T_{x}=T_{y}=3$ and $T_{0}=-2.5$ with two Dirac
points.}
\end{figure}

Another interesting situation is obtained from Eq.~(\ref{eq:disp_2D})
with $\phi_{x}=0$, $\phi_{y}=\pi/2$ and $T_{0}\neq0$ . Then, the
dispersion relation is, for small $k_{y}$
\begin{equation}
\omega(k_{x},k_{y})\approx\pm\sqrt{\left[T_{0}+T_{x}\cos(2k_{x})\right]^{2}+T_{y}^{2}k_{y}^{2}}\label{eq:Two_Dirac}
\end{equation}
Depending on parameters $T_{0}$ and $T_{x}$, different structures
are obtained (assuming here $T_{0}$ and $T_{x}$ of opposite sign
without loss of generality). As shown in Fig.~\ref{fig:2D-Dirac},
when $\left|T_{0}\right|<\left|T_{x}\right|$, there are two Dirac
cones centered at the positions $k_{y}=0$ and $k_{x}=\pm(1/2)\cos^{-1}\left(\left|T_{0}/T_{x}\right|\right)$.
These cones moves towards each other as the value of $\left|T_{0}/T_{x}\right|$
increases, and finally coalesce when $T_{0}=-T_{x}$ giving a dispersion
relation $\omega(k_{x},k_{y})\approx\pm\sqrt{4T_{x}^{2}k_{x}^{4}+T_{y}^{2}k_{y}^{2}}$
in the vicinity of the point $k_{x},k_{y}=0$, which is thus a hybrid
point with a linear dependence in the $y$ direction (corresponding
to a free Dirac particle, or phonon) and quadratic dependence in $x$
(corresponding to a free non-relativistic particle). A gap opens for
$\left|T_{0}\right|>\left|T_{x}\right|$, potentially leading to topological
effects. 

\subsection{The Lieb lattice}

\begin{figure}
\begin{centering}
\includegraphics[width=6cm]{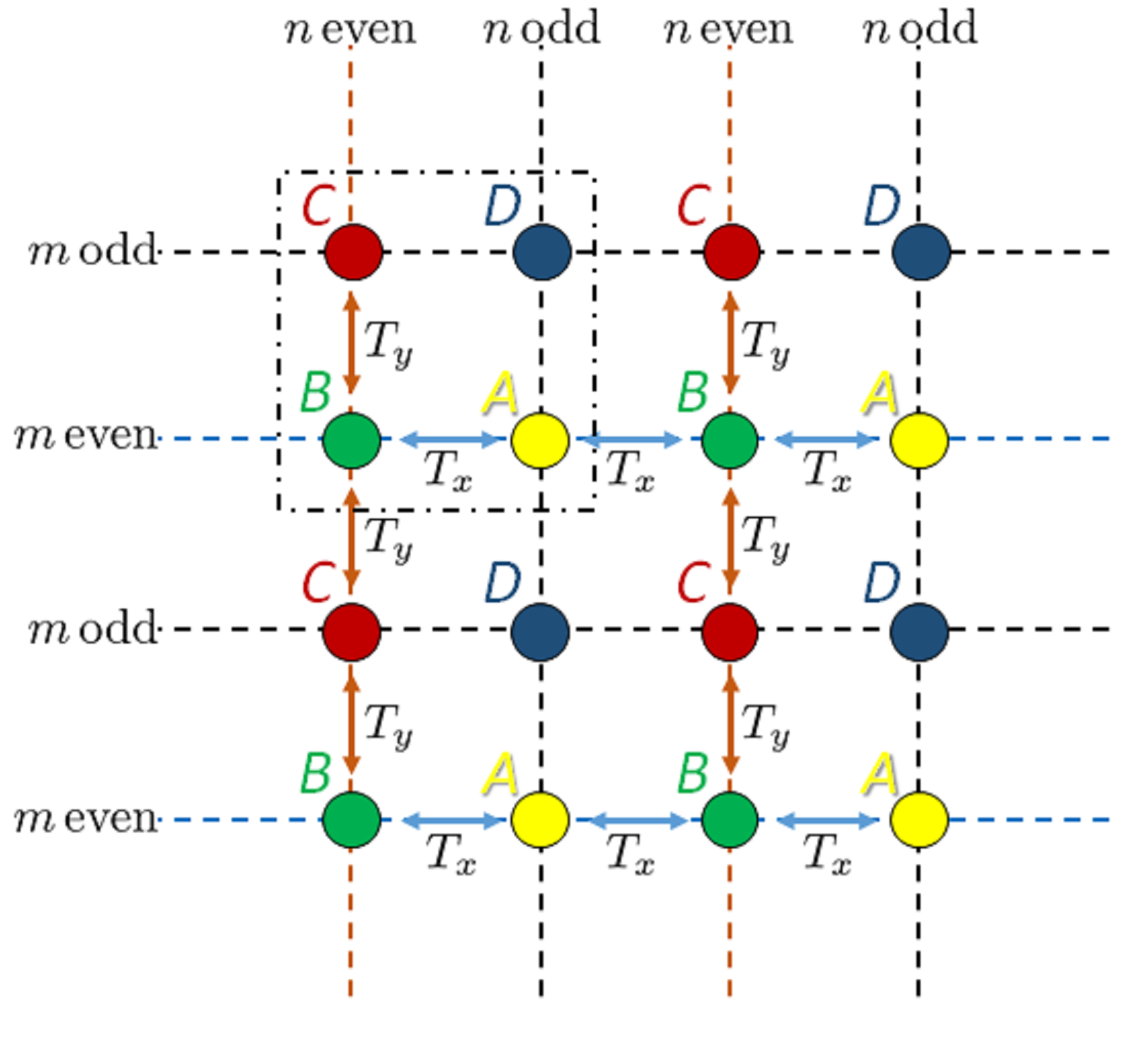}
\par\end{centering}
\caption{\label{fig:Lieb}Synthesis of a Lieb lattice. The unit cell is indicated
by the dashed-dotted square. Sites $D$ ($n$ and $m$ odd) are uncoupled
to their neighbors. The other sites $A,$$B$ and $C$ are resonantly
coupled to their neighbors: $B$ to $C$ with amplitude $T_{y}$ along
the lines of $m$ even; $A$ to $B$ with amplitude $T_{x}$ along
the lines of $n$ even. }
\end{figure}

In this section, we consider the synthesis of a Lieb lattice dispersion
relation~\citep{Slot:ExperimentalRealizationLiebLattice:NP17,Goldman:TopologicalPhasesLieb:PRA11,Flannigan:HubbardModelsLiebLattice:arXiv21},
using a form of Eq.~(\ref{eq:perturbation}) which results in an
effective spin $S=1$ with a flat band. The idea is sketched in Fig.~\ref{fig:Lieb}:
it essentially consists in creating different types of sites: $A,$
$B$, coupled with coupling $T_{x}$, $B$ and $C$ coupled with $T_{y}$,
while sites $D$ are completely uncoupled and thus dynamically irrelevant.

This selective coupling can be obtained with the following perturbation\begin{widetext}

\begin{align*}
H_{1_{2D}}(x,y,t)=\cos\left(2\pi x\right) & \left[V_{x}^{(1)}+V_{x}^{(2)}\cos(\pi y)\right]\cos\left(\omega_{B}^{(x)}t\right)+\cos\left(2\pi y\right)\left[V_{y}^{(1)}+V_{y}^{(2)}\cos(\pi x)\right]\cos\left(\omega_{B}^{(y)}t\right).
\end{align*}
\end{widetext}Substituting the general solution of Eq.~(\ref{eq:psi_2D})
in Schr\"odinger's equation for the Hamiltonian $H_{0_{2D}}+H_{1_{2D}}$
one obtains the following set of coupled equation (in the resonant
approximation)
\begin{align}
\im\frac{dc_{n,m}}{dt}= & T_{x}\left[1+(-1)^{m}\right]\left(c_{n+1,m}+c_{n-1,m}\right)\nonumber \\
 & +T_{y}\left[1+(-1)^{n}\right]\left(c_{n,m+1}+c_{n,m-1}\right)\label{eq:Lieb}
\end{align}
with couplings
\begin{align*}
T_{x} & =\frac{1}{2}V_{x}^{(1)}\left\langle \varphi_{0}^{(x)}\right|\cos\left(2\pi x\right)\left|\varphi_{1}^{(x)}\right\rangle \\
T_{y} & =\frac{1}{2}V_{y}^{(1)}\left\langle \varphi_{0}^{(y)}\right|\cos\left(2\pi y\right)\left|\varphi_{1}^{(y)}\right\rangle 
\end{align*}
provided that the modulation amplitudes obey the following relations
\begin{eqnarray*}
V_{x}^{(1)} & = & V_{x}^{(2)}\left\langle \varphi_{0}^{(y)}\right|\cos\left(\pi y\right)\left|\varphi^{(y)}\right\rangle \\
V_{y}^{(1)} & = & V_{y}^{(2)}\left\langle \varphi_{0}^{(x)}\right|\cos\left(\pi x\right)\left|\varphi^{(x)}\right\rangle .
\end{eqnarray*}
As expected, Eqs.~(\ref{eq:Lieb}) show that the complex amplitudes
$c_{n,m}$ with $n$ and $m$ odd (corresponding to $D$ sites) are
dynamically inert. These amplitudes are therefore irrelevant and,
as shown in Fig.~\ref{fig:Lieb}, the effective unit cell contains
only three relevant sites, which is the characteristic of Lieb lattices.

\begin{figure}
\begin{centering}
\includegraphics[width=6cm]{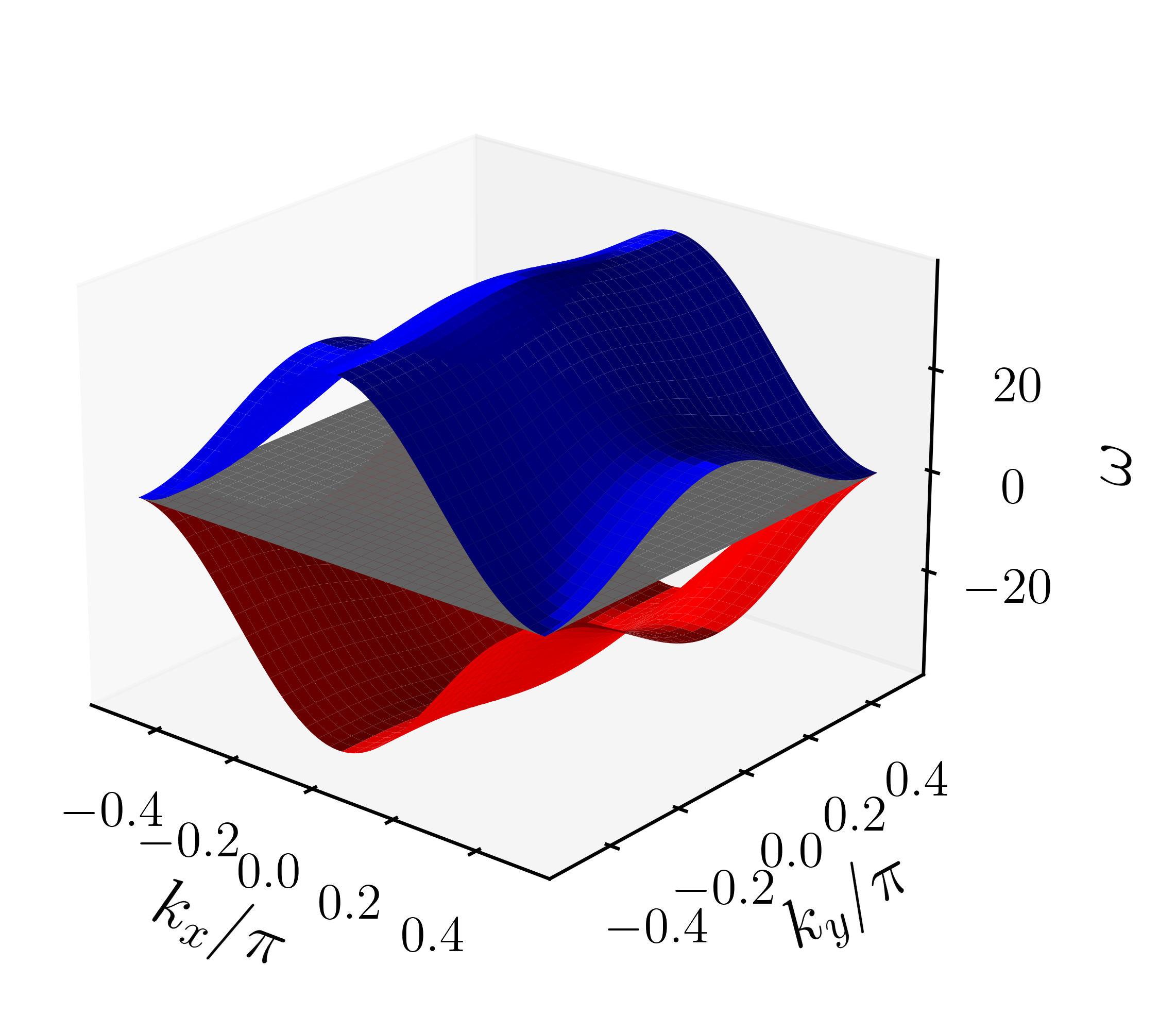}
\par\end{centering}
\caption{\label{fig:LiebLatticDisprel}Lieb lattice dispersion relation, Eq.~(\ref{eq:LiebDispRel}).
Parameters are $T_{x}=1.5$, $T_{y}=1$.}

\end{figure}

In the same spirit as in Sec.~\ref{subsec:2DDiracCones} we define
three amplitudes in $k$-space
\[
\tilde{c}(k_{x},k_{y},t)=\sum_{n\textrm{ even }}\sum_{m\textrm{ even }}c_{nm}(t)\mathrm{e}^{-\im nk_{x}}\mathrm{e}^{-\im mk_{y}},
\]
and $\tilde{d}(k_{x},k_{y},t)$ for $m$ even, $n$ odd, and $\tilde{f}(k_{x},k_{y},t)$
for $m$ odd, $n$ even. The time-evolution in $k$-space for the
three-component spinor $\left[\psi\right]=\left(\tilde{c}(k_{x},k_{y},t),\tilde{d}(k_{x},k_{y},t),\tilde{f}(k_{x},k_{y},t)\right){}^{\mathsf{T}}$
obeys
\[
\im\frac{d\left[\psi\right]}{dt}=\left[\begin{array}{ccc}
0 & 4T_{y}\cos k_{y} & 4T_{x}\cos k_{x}\\
4T_{y}\cos k_{y} & 0 & 0\\
4T_{x}\cos k_{x} & 0 & 0
\end{array}\right]\left[\psi\right]
\]
leading to the well-known three-band dispersion relations of the Lieb
lattice:
\begin{eqnarray}
\omega(k_{x},k_{y}) & = & 0\nonumber \\
\omega(k_{x},k_{y}) & = & \pm\sqrt{\left(4T_{x}\cos k_{x}\right)^{2}+\left(4T_{y}\cos k_{y}\right)^{2}}\label{eq:LiebDispRel}
\end{eqnarray}
where the first relation corresponds to the flat band and the second
one to two symmetric bands, and is anisotropic if $T_{x}\neq T_{y}$,
as shown in Fig.~\ref{fig:LiebLatticDisprel}.

\section{Conclusion}

We introduced in this work a technique allowing the generation of
arbitrary dispersion relations in a tilted modulated lattice, which
we illustrated through several important examples: the 1D Dirac phononic
dispersion relation, the Bogoliubov dispersion relation, the Landau
superfluid relation, with the maxon and the roton features. We proposed
a simple way to experimentally detect the dispersion relation by adding
a slow chirp to the modulation and measuring an easily accessible
quantity, namely the average position of the wave packet. Finally,
we illustrated a generation of Dirac points in 2 dimensions and the
generation of a flat band in a Lieb lattice. The technique introduced
in the present work thus appears as an efficient, state-of-the-art
experimentally feasible, way to synthesize lattice systems with arbitrary
dispersion relations, thus mimic a large variety of important condensed
matter systems.

\medskip{}

\begin{acknowledgments}
This work was supported by Agence Nationale de la Recherche through
MANYLOK project (Grant No. ANR-18-CE30-0017), the Labex CEMPI (Grant
No. ANR-11-LABX-0007-01), the Ministry of Higher Education and Research,
Hauts-de-France Council and European Regional Development Fund (ERDF)
through the Contrat de Projets \'Etat-R\'egion (CPER Photonics for
Society, P4S). 
\end{acknowledgments}

\appendix

\section{Dynamics in a chirped lattice}

\label{app:Appendix chirp}

In this appendix we show how the introduction of a chirp allows to
extract the shape of the dispersion relation from the measurement
of the average wave packet position evolution. 

Consider a smooth wave packet in a system obeying a dispersion relation
$\omega(k,t)$. Its average position $\left\langle X\right\rangle (t)$
can be written as 
\[
\left\langle X\right\rangle (t)=\left\langle X\right\rangle (t=0)+\int_{0}^{t}v_{G}(t^{\prime})dt^{\prime}
\]
where $v_{G}(t)=d\omega/dk$ is the group velocity.

An adiabatic chirp $\varphi(t)=\gamma t$ corresponds to $k(t)=k_{0}+\gamma t$,
and thus, by integration
\[
\left\langle X\right\rangle (t)=\left\langle X\right\rangle (t=0)+\left[\omega\left(k(t)\right)-\omega\left(k_{0}\right)\right]/\gamma
\]
with $\left\langle X\right\rangle (t=0)=0$. Choosing $k_{0}$ such
that $\omega(k_{0})=0$ this results in the simple expression: 
\[
\left\langle X\right\rangle (t)=\omega\left(k(t)\right)/\gamma.
\]
Hence, measuring of the temporal evolution of the average position
of the wave packet directly gives the shape of dispersion relation.


%

\end{document}